\documentclass[12pt]{article}

\usepackage{amssymb}
\def\beq{\begin{equation}}
\def\eeq{\end{equation}}
\def\beqn{\begin{eqnarray}}
\def\eeqn{\end{eqnarray}}
\def\={~=~}
\newcommand{\calle}[1]{(\ref{#1})}
\newcommand{\lp}{\left(}
\newcommand{\lb}{\left\lbrack}
\newcommand{\rp}{\right)}
\newcommand{\rb}{\right\rbrack}

\begin{document}

\rightline{\Large Finding Exact Values For Infinite Sums}

\vspace{5mm}
\rightline{COSTAS EFTHIMIOU}
\rightline{Department of Physics}
\rightline{Tel Aviv University}
\rightline{Tel Aviv, 69978 Israel}

\vspace{5mm}

{\bf  Introduction} In the 1995 February issue of {\it Math Horizons} 
I. Fisher posed the following problem
[1]:

\begin{minipage}{5in}
{\it From the well-known results,
$$
\sum_{n=1}^{\infty}\,{1\over n^2}={\pi^2\over 6}~~~~~~
{\rm and}~~~~~~
\sum_{n=1}^{+\infty}\,{1\over n^2+n}=1~,
$$
it follows that
$$
1<\sum_{n=1}^{\infty}\,{1\over n^2+n/2}<{\pi^2\over 6}~.
$$
Find the exact value of the convergent sum.}
\end{minipage}

This paper offers a solution method that  allows one to find exact values
for a large class of convergent series of rational terms.

In the next section, we first illustrate the method for a special case. 
We then
describe the  general result 
pointing out further generalizations of the method  
and we finally end with a brief discussion.

{\bf A Special Case} Consider the series
\beq
\label{sum}
  S(a,b)\=\sum_{n=1}^{\infty}\,{1\over (n+a)(n+b)}~,
\eeq
where $a\ne b$ and neither $a$ nor $b$ is a negative integer.
Sums of this form arise  often in 
problems dealing with Quantum Field Theory  (p. 89ff., Ref. [2]).

Decomposing each term of  (\ref{sum})  in 
partial fractions gives
\beq
\label{partialfract}
  S(a,b)\={1\over a-b}\,\sum_{n=1}^{\infty}\,
  \left( {1\over n+b}-{1\over n+a}\right)~.
\eeq

Now we  use  the identity
\beq
\label{laplace1}
   {1\over A}\=\int_0^{\infty}\,e^{-Ax}\, dx~,~~~~~A>0~.
\eeq
Therefore for $a,b>-1$ we have
\beqn
  S(a,b)&\=&{1\over a-b}\,\lim_{N\to+\infty}\sum_{n=1}^{N}\,
   \int_0^{\infty}\,e^{-nx}\, \left( e^{-bx}
  - \,e^{-ax}
  \right) \, dx 
  \nonumber\\
  &\=&
   \lim_{N\to+\infty}\int_0^{\infty}\, {e^{-bx} -e^{-ax}\over a-b}
  \,{e^{-x}\, (1-e^{-Nx})\over 1-e^{-x} }  \, dx
  \nonumber\\
  &\=&{1\over a-b}\,
   \int_0^{\infty}\, {e^{-bx} -e^{-ax}\over a-b}
  \,{e^{-x}\over 1-e^{-x} } \, dx ~.
\label{eq1}
\eeqn
In deriving the last result, we made use
of the monotone convergence theorem
(p.  318, Ref. [3]). In particular, 
 the integrand in the second line of equation \calle{eq1}
consists of a  non-decreasing sequence of non-negative
functions and therefore we can swap
the indicated operations of taking the limit and
performing the integration.

Making the change of variable $t=e^{-x}$ in the integral of
equation \calle{eq1}
gives a more symmetric result:
\beq
\label{generalresult}
  S(a,b)\={1\over a-b}\,\int_0^1\,{t^b-t^a\over 1-t}\, dt
   ~.
\eeq

Some comments are in order here:
\begin{itemize}

\item
Although the integral
$$
  \int_0^1\,{t^y\over 1-t}\, dt
$$
diverges, 
the integral of equation \calle{generalresult} converges
for $a\not= b$ and $a,b>-1$.

\item
The problem of Fisher that  appeared  in {\it Math Horizons} corresponds
to $a=1/2,~b=0$. In this case, we have 
$$
  \sum_{n=1}^{\infty}\,{1\over n(n+1/2)}
\=2\,\int_0^1\,{dt\over 1+\sqrt{t}}~.
$$
After the change of variables $t=u^2$, it is easy to  calculate the
integral:
$$
  \sum_{n=1}^{\infty}\,{1\over n(n+1/2)}
\=4\,\int_0^1\,\left( 1-{1\over 1+u}\right)\,
  du \=4(1-\ln2)\simeq 1.227 ~.
$$

\item
When the two numbers $a$ and $b$ differ by an integer $k$, i.e.
$a=b+k$, then the sum \calle{sum} ``telescopes" and it can 
be easily calculated from 
\calle{partialfract}:
$$
   S(a, a-k) \={1\over k}\, \sum_{j=1}^k\, {1\over j+a}~.
$$
This can be used as a consistency check of formula
\calle{generalresult}. Indeed
\beqn
  S(a, a-k) &=&  {1\over k}\, \int_0^1\,
   {t^k-1\over t-1}\, t^a \, dt 
   =  {1\over k}\, \int_0^1\,
    \sum_{i=0}^{k-1}\, t^{i+a} \, dt 
   =  {1\over k}\, \sum_{i=0}^{k-1}\,
    {t^{i+a+1}\over i+a+1}\Bigg|_0^1   
  \nonumber \\
   &=&  {1\over k}\, \sum_{j=1}^k\, 
    {1\over j+a} ~,
   \nonumber
\eeqn
in agreement with the last result. 

\end{itemize}
 
Now we note that one can
express the result \calle{generalresult}
in another equivalent  form ---
namely, using the well known representation (see p. 258, Ref. [4]) of the digamma
function $\psi(z)$:
\beq
\label{digamma}
   \psi(z)~\equiv~{d\over dz}\,\ln\Gamma(z)\=
 -\gamma-\int_0^1\,{t^{z-1}-1\over 1-t}\, dt~,
\eeq
where $\gamma$ is the Euler-Mascheroni constant
and $\Gamma(z)$ is the gamma function. This is motivated
by the fact that part of the integrand in the r.h.s. of equation
\calle{digamma} is  similar to the integrand in
\calle{generalresult}.
In fact, we have
\beq
\label{psi1} 
  S(a,b)\={\psi(b+1)-\psi(a+1)\over b-a} 
  ~.
\eeq

There are many useful identities involving the digamma function
(p. 258, Ref. [4]). For example
\beq
\label{relation1}
\psi(1+z)=\psi(z)+{1\over z}~.
\eeq
Moreover, the exact value of $\psi(z)$ is known
for several values of $z$:
\beq
\label{valuespsi}
   \psi(1)=-\gamma~,~~~~~
\psi(1/2)=-\gamma-2\ln2~.
\eeq

Equations \calle{relation1} and \calle{valuespsi} can be used
to evaluate $S(a,b)$ exactly for many values of
$a$ and $b$.
For example, if $a=1/2,~b=0$ we find
\beqn
  S(0,1/2)&\=& 2\, \lbrack \psi(3/2) -\psi(1)\rbrack
  \= 2\, \lbrack \psi(1/2)+2 -\psi(1)\rbrack
  \nonumber\\
  &\=& 2\, \lbrack -\gamma-2\ln2+2 +\gamma\rbrack
  \= 4\, (1-\ln2)~,
  \nonumber
\eeqn 
in agreement with our previous result.

When $a=b$ in \calle{sum}, the sum can still be calculated. 
We consider two approaches.
The first approach is to repeat the calculations presented above
but  observing that the basic equation (\ref{laplace1})
has now to be modified in the form
$$
  {1\over A^2} \= \int_0^\infty\, x\,e^{-Ax}\, dx~,~~~~~A>0.
$$
Following the same reasoning, we find
\beq
\label{eq:a=b}
  \sum_{n=1}^\infty\, {1\over (n+a)^2} \= -\int_0^1
  \, {t^a\, \ln t\over 1-t}\,dt~.
\eeq
Alternatively, we can obtain the same result by taking the limit
$b\rightarrow a$ in
(\ref{generalresult}):
\beqn
 \sum_{n=1}^\infty\, {1\over (n+a)^2} &=&\lim_{b\rightarrow a}
  \,\int_0^1\,{t^b-t^a\over b-a}
    \,{-1\over 1-t}\, dt ~.
\label{eq:1}
\eeqn
Without loss of generality, we can assume  that $-1<a<b$.
We notice that for $0\le t \le 1/2$
$$
 \left| {t^b-t^a\over 1-t}\,
         \right| \le 2\, t^b~,
$$
while for $1/2 < t \le 1$ the integrand 
of \calle{generalresult} is bounded; let $M$ be its
 supremum  in this subdomain. The function
$$
   g(t) =\cases{ {2\,t^b\over b-a}~, & if $~0\le t\le 1/2~,$\cr
                 M~, & if $~1/2< t\le 1~,$\cr}
$$
is integrable in $[0,1]$ and therefore the dominated convergence
theorem (p. 167, 321, Ref. [3]) can be used to interchange
the operations of the integral and  the limit in \calle{eq:1}:
\beqn
 \sum_{n=1}^\infty\, {1\over (n+a)^2} &=&
  \int_0^1\, \lim_{b\rightarrow a}{t^b-t^a\over b-a}
  \,{-1\over 1-t}\, dt
  =
  - \int_0^1\,
    \,{ t^a\, \ln t\over 1-t}\, dt~.
  \nonumber
\eeqn

Also, from equation \calle{psi1} we find
$$
   \sum_{n=1}^\infty\, {1\over (n+a)^2}\=
   {d\psi(z)\over dz}\Bigg|_{z=a+1}~.
$$
The functions
\beq
\label{polygamma}
   \psi^{(n)}(z)~\equiv~{d^{n+1}\over dz^{n+1}}\,\ln\Gamma(z)\=
 -\int_0^1\,dt\,{t^{z-1}\, (\ln t)^n \over 1-t}~,~~~~~n=1,2,\dots
\eeq
are known as polygamma functions (p. 260, Ref. [4]).
Several identities for the polygamma functions
are known (p. 258ff., Ref. [4]). For example
\beqn
   \psi^{(n)}(1)&=&(-1)^{n+1}\, n!\, \zeta(n+1)~, \nonumber \\
   \psi^{(n)}(1/2)&=&(-1)^{n+1}\, n!\,(2^{n+1}-1)\, \zeta(n+1)~,\nonumber \\
   \psi^{(n)}(z+1)&=&
   \psi^{(n)}(z)+ {(-1)^n\, n!\over z^{n+1}}~,\nonumber
\eeqn
where $\zeta(z)$ is the zeta function:
$$
   \zeta(z)\= \sum_{n=1}^\infty \, {1\over n^z}~,~~~~~
   {\rm Re}z>1~.
$$

As an application of \calle{eq:a=b}, we obtain the well known result
$$
 \sum_{n=1}^\infty\, {1\over n^2} \=
  - \int_0^1\,
    \,{  \ln t\over 1-t}\, dt 
  \= {\pi^2\over 6} ~\simeq~ 1.645 ~.
$$
Formula \calle{eq:a=b} also implies the less well known result
$$
 \sum_{n=1}^\infty\, {1\over (n+1/2)^2} \=
  - \int_0^1\,
    \,{\sqrt{t}\,  \ln t\over 1-t}\, dt \= 
 3\,\zeta(2)-4\={\pi^2\over 2} -4 ~=~ 0.935 ~.
$$

{\bf The General Case} After our preceding discussion,
we can now establish a more general result. Let
$$
  S\=\sum_{n=1}^{\infty}\,{Q_{N-2}(n)\over P_N(n)}~,
$$
where $Q_{N-2}(n), P_N(n)$ are two polynomials in $n$ of degree
$N-2$ and $N$ respectively. 
We shall assume that
$P_N(n)$ is expressible in the form
$$
  P_N(n)= (n+a_1)^{m_1}(n+a_2)^{m_2}\dots (n+a_k)^{m_k}~,
$$
with all $a_i,~i=1,2,\dots,k$ distinct real numbers none of which is
a negative integer.
This ensures the convergence
of $S$.
Then for any polynomial
$$
  Q_{N-2}(n)\= c_{N-2} \, n^{N-2}+ c_{N-3}\, n^{N-3}+\dots + c_0~,
$$
the sum
$S$ is written in terms of partial fractions:
$$
  S(a_1,\dots a_k; c_0,\dots,c_{N-2})
  \=\sum_{n=1}^\infty\sum_{i=1}^k\sum_{j=1}^{m_i}\,
           {A_{ij}\over (n+a_i)^j}~,
$$
where the constants $A_{ij}$ are uniquely determined by the
partial fraction decomposition of each summand. In particular,
notice that since there is no term of degree $N-1$ in $Q_{N-2}(n)$,
\beq
\label{A-constraint}
  \sum_{i=1}^k\,
           A_{i1}\=0~.
\eeq

Using the identity
\beq
\label{laplace2}
   {1\over A^L}\={1\over (L-1)!}\,
   \int_0^\infty\, x^{L-1} e^{-Ax}\, dx~,
\eeq
we write the series in an integral form 
valid only if  $a_i>-1,~\forall i$:
\beqn
  S(a_1,\dots a_k; c_0,\dots,c_{N-2})
  &=&\sum_{n=1}^\infty \sum_{i=1}^k \sum_{j=1}^{m_i}\,
  {A_{ij}\over (j-1)!}\, \int_0^\infty\,
    x^{j-1}\, e^{-(n+a_i)x}\, dx ~.
\eeqn
Working in a similar fashion
as in the derivation of equation \calle{eq1}, we find
\beqn
  S(a_1,\dots a_k; c_0,\dots,c_{N-2})
  &=&\sum_{i=1}^k \sum_{j=2}^{m_i}\,
  {A_{ij}\over (j-1)!}\, 
   \int_0^\infty\,
    x^{j-1}\, {e^{-(a_i+1)x}\over 1- e^{-x} }\,
    \, dx
  \nonumber \\
  &+&  \sum_{i=1}^k \,
  A_{i1}\, 
   \int_0^\infty\,
    {e^{-(a_i+1)x}-1\over 1- e^{-x} }\,
    \, dx~,
   \nonumber
\eeqn
where we have taken extra care for the $j=1$ term
(by using the condition \calle{A-constraint})
in order  to guarantee the convergence of the corresponding
integral.
This is our result in an integral form. We can also express it
in terms of the polygamma functions \calle{polygamma}: 
\beq
\label{gen-res}
  S(a_1,\dots a_k; c_0,\dots,c_{N-2})
  \=\sum_{i=1}^k \sum_{j=1}^{m_i}\,
  {(-1)^j\over (j-1)!}\,A_{ij}\, \psi^{(j-1)}(a_i+1)~,
\eeq
where  we have
defined $\psi^{(0)}(z)\equiv\psi(z)$.

As a straightforward application of our method, let us consider the 
following examples:

\begin{description}

\item \underbar{Example 1}:
$$
  S(a,-a)\=\sum_{n=1}^{+\infty}\,
     {1\over n^2-a^2}~,
$$
where $a$ is not a positive integer and satisfies the inequality $a>-1$.
Using the formula \calle{gen-res} we find
$$
  S(a,-a)\= {\psi(a+1)-\psi(-a+1) \over 2a}~,~~~~~a\ne 0~.
$$
This result can be further simplified if we make use of the functional
relation
$$
  \psi(-z+1)=\psi(z)+\pi\,\cot(\pi z)~,
$$
in conjunction with \calle{relation1}. Then
$$
  S(a,-a)\= {1 \over 2a}\,
  \lb {1\over a}-\pi\,\cot(\pi a) \rb~.
$$

\item \underbar{Example 2}:
$$
  S(\underbrace{a,\dots,a}_N)\=\sum_{n=1}^{+\infty}\,
     {1\over (n+a)^N}\=
{ (-1)^N\over (N-1)!}\, \psi^{(N-1)}(a+1)~,
$$
where $N\ge 2$ and $a>-1$.

\item \underbar{Example 3}:
$$
  S(0,0,3/2)\=\sum_{n=1}^{+\infty}\,
     {1\over n^2(n+1/2)}~.
$$
Using formula \calle{gen-res}, we find
$$
  S(0,0,3/2)\= 4\,\psi(1)+2\,\psi^{(1)}(1)-4\,\psi(3/2)\=
   {\pi^2\over 3}-8\,(1-\ln 2)
   ~=~ 0.835 ~.
$$

\item \underbar{Example 4}:
$$
  S(1,1,1/2)\=\sum_{n=1}^{+\infty}\,
     {1\over (n+1)^2 (n+1/2)}~.
$$
Using the formula \calle{gen-res} we find
$$
  S(1,1,1/2)\= \lb 4\, \psi(2)
  -2\,\psi^{(1)}(2)
  -4\,\psi(3/2)\rb \=
   2\,(4\ln 2-1)-
   {\pi^2\over 3} ~=~ 0.255 ~.
$$

\end{description}

Finally, the reader is invited to write down the values for other 
infinite sums
of the form presented above.

{\bf Discussion}
Before we finish, we would like to point out that the identities
\calle{laplace1} and \calle{laplace2} we used
in our derivations express the quantities
$1/A$ and  $1/A^L$  as the
Laplace transforms of $1$ and $x^{L-1}$
 respectively.
In general, if $f(s)$ is the Laplace transform of $g(x)$,
$$
    f(s)\=\int_0^\infty\,e^{-sx}\,g(x)\, dx~,
$$
then the sum
$$
    S_I\= \sum_{n\in I}\, f(n)~,
$$
where $I\subset{\mathbb{Z}}$, 
can be written in the form
$$
    S_I\= \int_0^\infty\, g(x)\, \lp \sum_{n\in I}\,e^{-nx}\rp
    \, dx ~,
$$
assuming that the operations of summation and integration are 
interchangable.
Assuming moreover
that the sum inside the parenthesis can be performed explicitly, we
have thus obtained an integral representation of $S_I$.

The Laplace transform has been proved a very valuable tool in the solution
of differential equations. Unfortunately, in the summation of series,
the Laplace transform does not enjoy the same popularity. In this paper,
we have tried to present some of the
 limitless possibilities that the method offers.
We propose our reader  to solve the following problem:

\underbar{Problem}

{\it
 (i) Show that
$$
 \sum_{n=1}^\infty\, {(-1)^{n+1}\over n+a}\=\int_0^1\, {t^a\over 1+t}
 \, dt~.
$$

(ii) Using the previous result, derive the well known result    
$$
 \sum_{n=1}^\infty\, {(-1)^{n+1}\over n}\=\ln2~=~ 0.693~,
$$
\indent and the less known result
$$
 \sum_{n=1}^\infty\, {(-1)^{n+1}\over n+1/2}\=2-{\pi\over 2}
  ~=~ 0.429~.
$$
}

We hope that this will   motivate him/her to explore more aspects of the method
presented in this paper and establish many additional results.

{\bf Acknowelegments}
The author would like to thank the referees of the paper 
for their valuable
comments and help during the revision of the initial version
of the paper. 
Also, he thanks the Cornell High Energy Group
where the preliminary version of this paper was written.

\vspace{5mm}
REFERENCES

\vspace{2mm}

\noindent [1]
I. Fischer, Problem 23 in Problem Section,
{\it Math Horizons}, February 1995.

\noindent [2]
P. Ramond,  {\it Field Theory: A Modern Primer},
Addison-Wesley, New York,  USA, 1994. 

\noindent [3]
W. Rudin, {\it Principles of Mathematical Analysis}, 3rd edition,
McGraw-Hill, Inc., New York, 1976.

\noindent [4]
M. Abramowitz and I. A. Stegun, {\it Handbook of Mathematical
Functions with Formulas, Graphs 
and Mathematical Tables}, Dover Publications,
New York, USA, 1972.

\end{document}